\documentclass[a4paper,english]{siamltex}
\pdfoutput=1
\usepackage[T1]{fontenc}
\usepackage[latin9]{inputenc}
\usepackage{textcomp}
\usepackage{amsmath}
\usepackage{graphicx}
\usepackage{amssymb}
\usepackage{listings}

\makeatletter

\providecommand{\tabularnewline}{\\}

\usepackage{hyperref}
\usepackage{booktabs}
\providecommand{\OO}[1]{\operatorname{O}\!\left(#1\right)}

\makeatother

\usepackage{babel}

\begin{document}

\title{The Piranha algebraic manipulator}

\author{Francesco Biscani\thanks{Research fellow at the Advanced Concepts Team, European Space Agency.
Previously at the Astronomy Department of the University of Padua
(\href{mailto:bluescarni@gmail.com}{bluescarni@gmail.com}).}}

\maketitle
\begin{keywords}
Celestial Mechanics, algebraic manipulation, computer algebra, Poisson
series, multivariate polynomials
\end{keywords}
\begin{abstract}
In this paper we present a specialised algebraic manipulation package
devoted to Celestial Mechanics. The system, called Piranha, is built on top
of a generic and extensible framework, which allows to treat efficiently and in a unified
way the algebraic structures most commonly encountered in Celestial Mechanics (such as multivariate polynomials
and Poisson series). In this contribution
we explain the architecture of the software, with special focus on the implementation of series arithmetics, show
its current capabilities, and present benchmarks indicating that Piranha is competitive, performance-wise, with other specialised
manipulators.
\end{abstract}

\section{Introduction}

Since the late Fifties (\cite{Herget59}) researchers in the field
of Celestial Mechanics have manifested a steady and constant interest
in software systems able to manipulate the long algebraic expressions
arising in the application of perturbative methods. Despite the widespread
availability of commercial general-purpose algebraic manipulators,
researchers have often preferred to develop and employ specialised
ad-hoc programs (we recall here, without claims of completeness, \cite{Broucke69},
\cite{Jefferys70,Jefferys72}, \cite{Rom70}, \cite{CAMAL}, \cite{UPP},
\cite{Dasenbrock82}, \cite{Parsec89}, \cite{Abad94}, \cite{Ivanova96},
\cite{Gregoire,Colbert} and \cite{TRIP2008}). The reason for this
preference lies mainly in the higher performance that can be obtained
by these. Specialised manipulators are built to
deal only with specific algebraic structures, and thus they can adopt
fast algorithm and data structures and avoid the overhead inherent
in general-purpose systems (which instead are designed to deal with
a wide variety of mathematical expressions). The performance gap between
specialised and general-purpose manipulators is often measured in
orders of magnitude, especially for the most computationally-intensive
operations.

The need for computer-assisted algebraic manipulation in Celestial
Mechanics typically arises in the context of perturbative methods.
In classical perturbative approaches, for instance, the equations
of motion of celestial bodies are expanded into Fourier series having
power series of ``small'' quantities (e.g., orbital eccentricity,
orbital inclination, ratio of the masses) as coefficients. This algebraic
structure, commonly referred to as Poisson series, allows to identify
the terms relevant to the specific physical problem being considered
and apply adequate methods of solution of the equations of motion
(\cite{SSD2000}, Chapters 7 and 8). Likewise, in the computation
of normal forms in studies of dynamical systems the Hamiltonian is
commonly expanded into power series, hence leading to multivariate
polynomial algebraic expressions (\cite{jorba98methodology}). Another
example is the computation of the tide-generating potential (\cite{DoodsonTGP,roosbeek95}),
which involves the manipulation of Fourier series with numerical coefficients.

In order to give a simple example, we recall here that classical perturbative
methods employ Fourier series expansions like this one (\cite{SSD2000},
Chapter 2),\begin{multline}
\cos f=\cos M+e\left(\cos2M-1\right)+\frac{9}{8}e^{2}\left(\cos3M-\cos M\right)\\
+\frac{4}{3}e^{3}\left(\cos4M-\cos2M\right)\\
+e^{4}\left(\frac{25}{192}\cos M-\frac{225}{128}\cos3M+\frac{625}{384}\cos5M\right),\end{multline}
which represents the expansion of the cosine of the true anomaly $f$
of the orbit of a celestial body in terms of cosines of the mean anomaly $M$ up
to fourth order in the eccentricity $e$. Such series are manipulated
and combined with other series through operations such as addition,
subtraction, multiplication, differentiation and symbolic substitution,
leading to a final Fourier series representing the perturbing potential.

All the mentioned applications require the ability to perform simple
operations on series whose number of elements can grow remarkably
large. In modern studies on the long-term stability of the Solar System,
for instance, researchers have to deal with series of millions of
terms (\cite{Kuzntesov2004PSP}). The fact that in many cases it is
possible to perform the most computationally-intensive calculations
within the form of a single algebraic structure might explain the
persisting interest in specific algebraic manipulators.

To improve performance with respect to general-purpose systems, different
specialised manipulators adopt similar techniques which are a natural
exploitation of the characteristics of the problems that they are
built to tackle. Exponents in polynomials, for instance, are usually
represented as hardware integers, instead of the arbitrary-size integers
employed in general-purpose systems. Perturbative methods, indeed,
assume that the quantities subject to exponentiation are small, so
that in practice the exponents encountered in such calculations have
values below $10^{2}$ (but usually they are lower). Similarly, when
calculating the harmonic expansion of the tide-generating potential
it is often sufficient to represent the series' coefficients as standard
double-precision floating point variables, whereas general-purpose
manipulators must be able to operate with multiprecision floating
point arithmetic and thus incur in an overhead that is unnecessary
from the point of view of the celestial mechanician.

The work presented here has four major objectives:

\begin{enumerate}
\item to define a framework able to handle concisely and in a generalised
fashion the algebraic structures most commonly encountered in Celestial
Mechanics,
\item to identify and implement efficient algorithms and data structures
for the manipulation of said algebraic structures,
\item to provide means to interact effectively and flexibly with the computational
objects, and
\item to provide an open system promoting participation and extensibility.
\end{enumerate}
We feel that many of the previous efforts in the field of algebraic
manipulators devoted to Celestial Mechanics have failed in at least
one of these objectives. Many manipulators, for instance, have been
written for very specific tasks, so that it can be difficult to extend
or reuse them in other contexts. Others employ algorithms and data
structures whose performance is not optimal. Yet others are fast and
full-featured, but they are not open.

While the work presented here is still in progress, preliminary benchmarks
seem to indicate that our system, called Piranha, is competitive performance-wise
with other more mature systems. In this contribution we illustrate
Piranha's architecture and its most noteworthy implementation details,
with special emphasis on series arithmetic. We also present a brief
overview of the Python interface and expound on benchmarks. Finally,
we discuss the future direction of the project, including desirable
features not yet implemented and possible performance improvements.

\section{Algebraic structures in Celestial Mechanics\label{sec:Manipulation-of-what}}

As hinted in the introduction, Celestial Mechanics employs diverse
algebraic structures in the form of series. Multivariate polynomials
appear naturally in perturbation theories in the form\begin{equation}
\sum_{\mathbf{i}}C_{\mathbf{i}}\mathbf{x}^{\mathbf{e}_{\mathbf{i}}},\label{eq:poly_01}\end{equation}
where we have adopted the convention\begin{equation}
\mathbf{x}^{\mathbf{e}_{\mathbf{i}}}=x_{0}^{e_{i_{0}}}x_{1}^{e_{i_{1}}}\ldots x_{n-1}^{e_{i_{n-1}}}\end{equation}
with $\mathbf{i}=\left(i_{0},i_{1},\ldots,i_{n-1}\right)$ an integer
multiindex, $\mathbf{e}_{\mathbf{i}}$ a vector of integer exponents
indexed over $\mathbf{i}$, $C_{\mathbf{i}}$ a numerical coefficient
indexed over $\mathbf{i}$ and $\mathbf{x}=\left(x_{0},x_{1},\ldots,x_{n-1}\right)$
a vector of symbolic variables. Although in perturbation theories
series expansions have infinite terms, in practical calculations series
are truncated to some finite order, so that the components of the
$\mathbf{i}$ vectors vary on finite ranges. Please note that, strictly
speaking, formula (\ref{eq:poly_01}) represents a superset of multivariate
polynomials, since the $\mathbf{e}_{\mathbf{i}}$'s components are allowed
to assume negative values. It is hence more correct to refer to expressions
in the form of (\ref{eq:poly_01}) as multivariate Laurent series,
as pointed out in \cite{San-Juan01}. Nevertheless, when referring
to polynomials from now on we will encompass also Laurent series.

In Celestial Mechanics another commonly-encountered algebraic structure
is the so-called Poisson series (\cite{Danby66})\begin{equation}
\sum_{\mathbf{i}}\sum_{\mathbf{j}}C_{\mathbf{ij}}\mathbf{x}^{\mathbf{e}_{\mathbf{i}\mathbf{j}}}\begin{array}{c}
\cos\\
\sin\end{array}\left(\mathbf{t}\cdot\mathbf{i}\right),\end{equation}
which consists of a multivariate Fourier series in the trigonometric
variables $\mathbf{t}$ with multivariate Laurent series as coefficients.
We refer to the elements of the integer vector $\mathbf{i}$ as \emph{trigonometric
multipliers}. In practical applications it is sometimes convenient
to represent the trigonometric parts of Poisson series through complex
exponentials, hence defining a structure which we refer to as \emph{polar}
Poisson series:\begin{equation}
\sum_{\mathbf{i}}\sum_{\mathbf{j}}C_{\mathbf{ij}}\mathbf{x}^{\mathbf{e}_{\mathbf{i}\mathbf{j}}}\exp\left[\imath\left(\mathbf{t}\cdot\mathbf{i}\right)\right],\end{equation}
where $\imath=\sqrt{-1}$. Poisson series arise routinely in the formulation
of the gravitational disturbing function, where the symbolic variables
represent the orbital elements of celestial bodies (\cite{SSD2000},
Chapter 6).

The subset of degenerate Poisson series with purely numerical coefficients
is often referred to simply as Fourier series:\begin{equation}
\sum_{\mathbf{i}}C_{\mathbf{i}}\begin{array}{c}
\cos\\
\sin\end{array}\left(\mathbf{t}\cdot\mathbf{i}\right).\end{equation}
Fourier series are used to express the periodic solutions of theories
of motion of celestial bodies. They are often employed in problems
involving the harmonic expansion of the tide-generating potential,
where the theories of motion of celestial bodies are plugged into
the expression of the potential and manipulated in order to be obtain
a formulation of the potential in Fourier series form.

Poisson series can be seen as a subset of \emph{echeloned} Poisson
series, which are defined by the following formula:\begin{equation}
\sum_{\mathbf{i}}\sum_{\mathbf{j}}\sum_{\mathbf{k}}C_{\mathbf{ijk}}\mathbf{x}^{\mathbf{e}_{\mathbf{i}\mathbf{j}\mathbf{k}}}\frac{1}{\prod_{\mathbf{l}}\left(\mathbf{d}\cdot\mathbf{m}_{\mathbf{i}\mathbf{j}\mathbf{l}}\right)^{\delta_{\mathbf{i}\mathbf{j}\mathbf{l}}}}\begin{array}{c}
\cos\\
\sin\end{array}\left(\mathbf{t}\cdot\mathbf{i}\right),\end{equation}
where $\mathbf{d}$ is a vector of symbolic variables (the so-called
\emph{divisors}), $\mathbf{m}$ a vector of integers indexed over
the multiindices $\mathbf{i}$, $\mathbf{j}$ and $\mathbf{l}$, and
$\delta_{\mathbf{i}\mathbf{j}\mathbf{l}}$ a positive integer value,
again indexed over $\mathbf{i}$, $\mathbf{j}$ and $\mathbf{l}$.
Echeloned Poisson series are employed, for instance, in lunar theories,
where it is necessary to express symbolically the frequencies of the
trigonometric variables. Such frequencies appear as divisors when
integrating the equations of motion with respect to time. While Poisson
series manipulators are relatively common, echeloned Poisson series
manipulators are less widespread (\cite{RomESP}, \cite{IvanovaEPSP}).

There are other less common algebraic structures for which specialised
algebaric manipulators exist. We recall here the mainpulator described
in \cite{Navarro2002}, which targets Earth rotation theories and
handles a superset of echeloned Poisson series called \emph{Kinoshita}
series.

In order to accomodate all the aforementioned algebraic structures,
Piranha adopts a definition of series based on the following concepts:

\begin{definition}
Terms are pairs constituted by a coefficient and a key.
\end{definition}

\begin{definition}
Two terms are considered equivalent if and only if their keys are.
\end{definition}

\begin{definition}
Series are sets of terms.
\end{definition}

\begin{figure}
\begin{centering}
\includegraphics[scale=0.5]{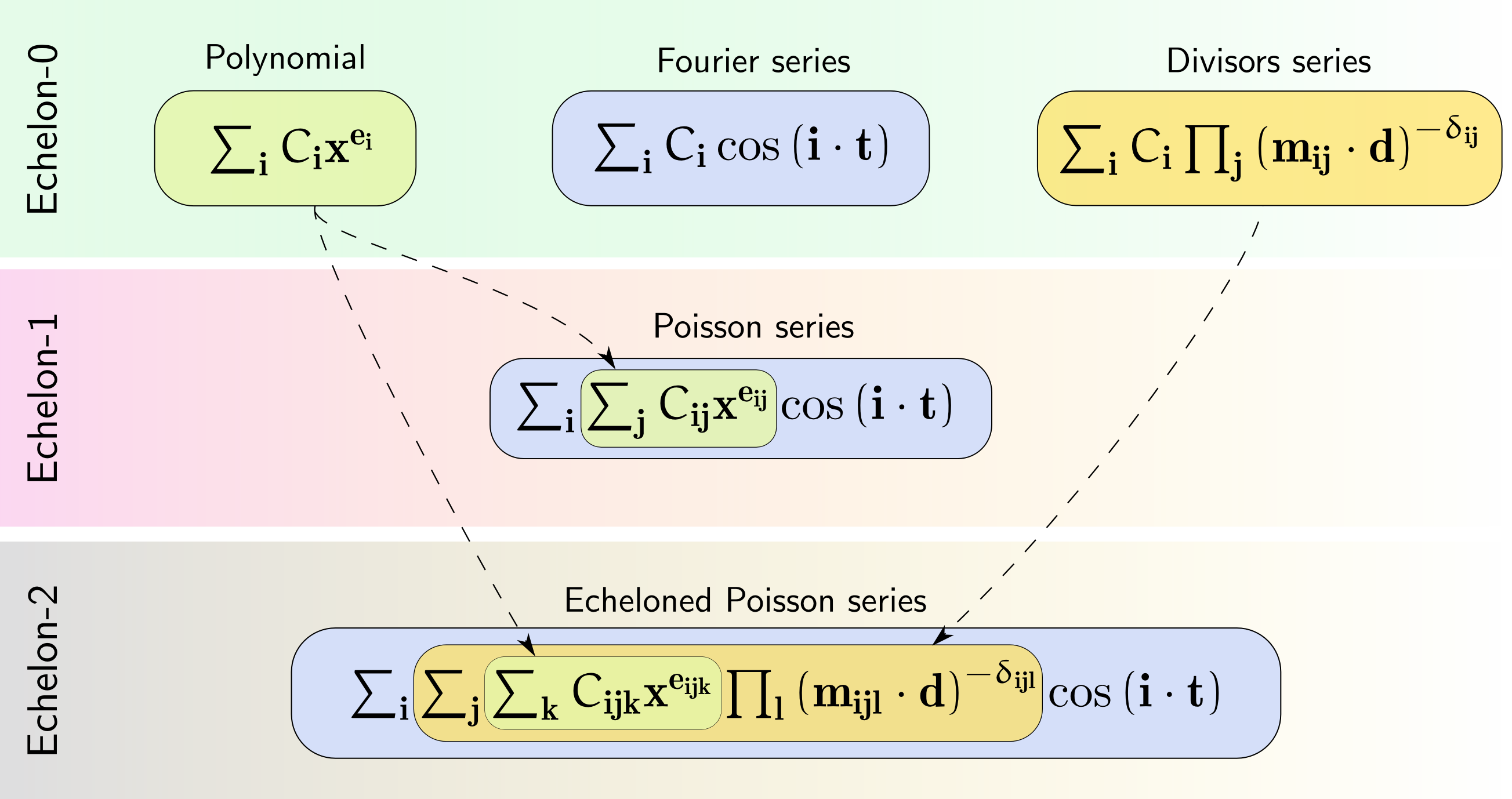}
\par\end{centering}

\caption{Definition of series in Piranha. Echelon-0 series are series with
numerical coefficients, echelon-$n$ series with $n>0$ are series
with echelon-$\left(n-1\right)$ series as coefficients.\label{fig:echelon_n}}

\end{figure}

Such concepts allow to introduce a recursive definition of series
which is exemplified in Figure \ref{fig:echelon_n}, and which was
inspired by the definition of echeloned Poisson series in \cite{RomESP}.
The lowest level of this scheme is populated by series whose coefficients
are numerical, such as Fourier series (where we find trigonometric
keys) and polynomials (where keys are multivariate monomials). This
level represents the terminal point of the recursion and we refer
to it as \emph{echelon}-0. The next level, echelon-1, features series
whose coefficients are echelon-0 series, and it is populated by Poisson
series (in both their traditional and polar forms). Similarly echelon-2,
featuring echeloned Poisson series, is characterised by series whose
coefficients are echelon-1 series. The following generalisation now
follows straightforwardly:

\begin{definition}
Echelon-0 series are series with numerical values
as coefficients.
\end{definition}

\begin{definition}
Echelon-$n$ series are series with echelon-$\left(n-1\right)$
series as coefficients.
\end{definition}

At the present time, Piranha implements Fourier series, polynomials
and Poisson series, with various possible combinations for the representation
of numerical coefficients and keys. In our opinion, one of the key
advantages of this framework is that it allows to treat different
algebraic structures in a unified way and without any overhead. While
it is possible, for instance, to use a Poisson series manipulator
to operate on multivariate polynomials and Fourier series, this approach
incurs in a performance penalty related to the unnecessary management
of empty trigonometric keys and monomials, respectively. By using the
framework described here, it is instead possible to build specialised
manipulators with little effort.

\section{A generically-programmed architecture}

In order to implement in an efficient and concise way the recursive scheme
described in \S\ref{sec:Manipulation-of-what}, we have chosen to
adopt a programming paradigm which is both generic and object-oriented.
The kind of genericity we seek is twofold:

\begin{enumerate}
\item we want to be able to implement series types with different representations
for coefficients and keys. When undertaking the task of the development
of the disturbing function in planetary problems, for instance, it
may be desirable to operate with multiprecision rational numbers as
coefficients, in order to retain maximum precision. Likewise, in certain
problems it may be enough to operate with polynomial exponents or
trigonometric multipliers represented as 16 or even 8 bits integers
as opposed to word-size integers, so that it is possible to adopt
a more compact representation for keys in such cases;
\item consequently, we want to be able to write algorithms for the manipulation
of terms once and for all the different combinations of coefficients
and keys, at the same time retaining the ability to specialise the
algorithms under certain circumstances.
\end{enumerate}
The ultimate goal is to minimise code duplication and keep the source
code as compact as possible. In a sense, we perceive Piranha as an attempt
to implement the algebraic manipulation framework envisioned in \cite{Henrard88}.

Our language of choice for the implementation of Piranha's architecture
is C++, in its flavour that is sometimes referred to as \emph{modern}
C++ (a term first popularised in \cite{ModernC++}). This programming
paradigm uses extensively C++ templates to achieve genericity at compile-time
and to provide metaprogramming capabilities (meaning that the program
is able to modify itself and automatically generate source code at
compile-time). Modern C++ can be used to provide container classes
where the type of the contained items is parametrised, to enable a
type of polymorphism which is entirely resolved at compile-time (in
contrast to the runtime polymorphism of traditional C++), and to write
algorithms able to operate on generic types. Since template metaprogramming
operates during the compilation of the program, its overhead at runtime
is null, and thus it is particularly appealing in those contexts,
such as scientific computing, where performance is paramount.

In Piranha, series are represented as template classes with coefficients
and keys as parametrised types. A \emph{base} series class provides
the bulk of low-level operations on series. Base series are not intended
to be used directly, instead they serve as a foundation for two other
series classes, i.e., \emph{named} series and \emph{coefficient} series.
Named series are intended to be employed directly by the user, and
they are so called because they embed a description of their arguments
in a $n$-tuple of vectors (where $n$ is the echelon level plus one).
The series types described in \S\ref{sec:Manipulation-of-what} are
all named series. Coefficient series, as the name suggests, are instead
series serving as coefficients in other series. Laurent series as
coefficients in Poisson series are an example of coefficient series.

All the series classes implemented in Piranha inherit from the base
series class and either from the named series class or the coefficient
series class. A minimal series class defined this way is able to read
and save series files, to interact with plain old C++ data types (i.e.,
integers and double-precision values), to produce a numerical evaluation
of itself based on the substitution of arguments with numerical values
and to perform series addition and subtraction. The capabilities of
the series are then augmented, again through multiple inheritance,
by the use of additional classes which we refer to as \emph{toolbox}
classes. Toolboxes are used to provide routines for more advanced
tasks such as series multiplication, exponentiation to real powers,
trigonometric functions, special functions, series expansions relevant
to Celestial Mechanics and so on, and also to group together methods
relevant only to specific series types. Toolboxes are also used to
provide those capabilities expected from series with complex coefficients
and which are useless in series with real coefficients, such as the
extraction of real and imaginary parts and full arithmetical interoperability
between complex and real series.

As an example, the Poisson series class at the present times inherits
from the following classes: base series, named series, series multiplication
toolbox, power series toolbox, special functions toolbox, Celestial Mechanics toolbox
and common Poisson series toolbox.

Toolboxes are generically programmed and may cross-reference each
other. The special functions toolbox, for instance, requires the presence
of a series multiplication toolbox (which might be already provided
by the framework, but which can also be (re)implemented by the user);
if the series multiplication toolbox is not available, a compile-time
error will be produced. To avoid runtime overhead, the system of toolboxes
extensively employs a design pattern known as \emph{curiously-recurring
template pattern} (\cite{CRTP95}). This approach avoids runtime
virtual-table lookups by achieving a form of polymorphism which is
resolved entirely at compile-time (and which has been sometimes referred
to as \emph{static} polymorphism).

At the time of this writing, Piranha supports the following coefficients
types:

\begin{enumerate}
\item standard double-precision floating-point,
\item multiprecision integers,
\item multiprecision rationals,
\end{enumerate}
and their complex counterparts. The multiprecision coefficients use
the GMP bignum library (\cite{GMP}). The supported key types are:

\begin{enumerate}
\item dynamically-sized trigonometric array (for Fourier and Poisson series),
\item dynamically-sized multivariate monomial array (for polynomials).
\end{enumerate}
For both keys the user may choose as integer size either 8 or 16 bits
at compile-time. Considering that Piranha currently supports three
named series types (polynomials and Poisson/Fourier series), there
are tens of possible manipulators implementable with a one-line type
definition. The following statement\footnote{The actual type definition in the source code is syntactically a bit
more verbose, but we chose to omit some details for better readability.}, for instance, defines a Fourier series class called \lstinline[basicstyle={\ttfamily}]!dfs!
with double-precision floating-point coefficients and 16 bits trigonometric
keys:
\begingroup
\inputencoding{latin1}
\begin{lstlisting}[basicstyle={\ttfamily},language={C++}]
typedef fourier_series<double_cf, trig_array<16> > dfs;
\end{lstlisting}
\endgroup

In order to evaluate the global impact of the framework, we have run
a simple SLOC (single line of code) analysis on Piranha and on two
freely-available specialised manipulators called Gregoire and Colbert
(\cite{Gregoire}, \cite{Colbert}) which target Fourier series (Gregoire)
and Poisson series (Colbert) with floating-point coefficients. Gregoire
and Colbert, whose feature set is comparable to Piranha's, are written
in Fortran and they consist of around $4\,000$ SLOCs each. By comparison
Piranha's SLOC count is about $10\,000$, but, as pointed out earlier,
Piranha really implements many manipulators with different characteristics
and different representations for coefficients and keys. Another interesting
detail of our simple SLOC analysis is that around 70\% of Piranha's
source code is shared among all the available series types.

\section{Storage of terms\label{sec:Storage-of-terms}}

Series appearing in the context of Celestial Mechanics are generally
sparse. By this we mean that, given finite upper and lower boundaries
for the multiindices indexing the series described in \S\ref{sec:Manipulation-of-what},
most of the possible multiindices combinations will be associated
to null terms. This is the main reason that led us to choose hashing
as storage method for series in Piranha.

We recall here briefly that hashing is an indexing technique based
on a \emph{hash function} that maps the items to be stored to integer
values (\emph{hash values}). Such values are used as indices in array-like
structures called \emph{hash tables} (sometimes called also \emph{dictionaries}).
The choice of the hash function, the strategy of collision resolution
(i.e., how to cope with different items hashing to the same value),
the resize policy, all contribute to the performance of the hash table
and they are thoroughly analysed in standard computer science textbooks
such as \cite{knuth98_3} and \cite{Cormen90}. We just recall
that, when properly implemented, hash tables feature expected $\OO{1}$
complexity for lookup, insertion and deletion of elements.

Given the definition of series provided in \S\ref{sec:Manipulation-of-what},
it readily follows that, since terms are uniquely identified by their
keys, the hash function will take the term's key as argument and that
coefficients have no role in the identifcation of a term. The presently-supported
key types (i.e., trigonometric keys and multivariate monomials) can
be seen as sequences of integers, with the flavour of trigonometric
keys (i.e., sine or cosine) codified as a boolean flag:\begin{eqnarray}
\cos\left(x+2y-3z\right) & \Longrightarrow & \left[1,2,-3,\textnormal{true}\right],\\
\sin\left(x-z\right) & \Longrightarrow & \left[1,0,-1,\textnormal{false}\right],\\
x^{3}y^{4}z^{-5} & \Longrightarrow & \left[3,4,-5\right].\end{eqnarray}
At the present time, Piranha implements the supported key types as
dynamically-sized dense arrays, meaning that the integer values are
stored in contiguous memory areas allocated dynamically and that null
values are explicitly stored. This type of representation
is referred to as \emph{distributed} in the literature, as opposed to the \emph{recursive} representation
in which $n$-variate series are recursively represented as $\left( n - 1 \right)$-variate series with univariate
series as coefficients (see \cite{Stoutemyer84} for a comparison of polynomial
representations in computer algebra systems).

Both trigonometric keys and monomials are implemented on top of
a base class called \lstinline[basicstyle={\ttfamily}]!int_array!,
which provides common low-level functionality. Since, as mentioned
earlier, we don't need the full range of word-size integers for the
representation of keys, \lstinline[basicstyle={\ttfamily}]!int_array!
employs an integer packing technique to store multiple sub-word-size
integers into a word-size integer. The user can choose a size for
the packed integers of either 8 or 16 bits at compile-time. The data
layout of a multivariate monomial in eight variables on a 64 bits
architecture with packed integer size of 16 bits will then look like
this:\[
\underbrace{\begin{array}{cccc}
\underbrace{n_{0}}_{16\textnormal{ bits}} & n_{1} & n_{2} & n_{3}\end{array}}_{64\textnormal{ bits}}\underbrace{\begin{array}{cccc}
\underbrace{n_{4}}_{16\textnormal{ bits}} & n_{5} & n_{6} & n_{7}\end{array}}_{64\textnormal{ bits}}\]
By using a C \lstinline[basicstyle={\ttfamily}]!union! it is possible
to acces the memory space as an array of either word-size or packed-size
integers. In addition to saving memory, this approach also allows
to increase the performance of those operations which can work on
multiple values independently; when comparing two keys, for instance,
we can compare directly the word-size integers, hence avoiding the
overhead of looping on the packed values and performing the comparison
of multiple packed integers in one pass. Similarly, it is possible
to compute the hash value of the key by combining (or \emph{mixing})
the word-size integers instead of the packed integers.

At the present time, Piranha relies on the hashing facilities provided
by the Boost C++ libraries (\cite{Boost}). Boost's \lstinline[basicstyle={\ttfamily}]!hash_combine!
function is based on a family of hash functions described in \cite{ramakrishna97},
initially conceived for string hashing. This class of hash functions
is refferred to as \emph{shift-add-xor} because at each step of the
iteration for the calculation of the hash value, operations of bit
shifting, addition and bitwise exclusive OR are used. In Piranha the
first word-size integer of each key is used as a seed value and then
mixed with the other word-size integers of the key. For trigonometric
keys, additional mixing is provided by the flavour of the key (cast
as an integer value). The hash value is then used to store the term
in a Boost \lstinline[basicstyle={\ttfamily}]!hash_set! data structure.
This hashing container is a fairly standard hash set which uses prime
numbers as sizes and the modulo operation for extracting a useful
index from the hash value, and which employs separate chaining for
collision resolution.

\begin{figure}
\begin{centering}
\includegraphics[scale=0.5]{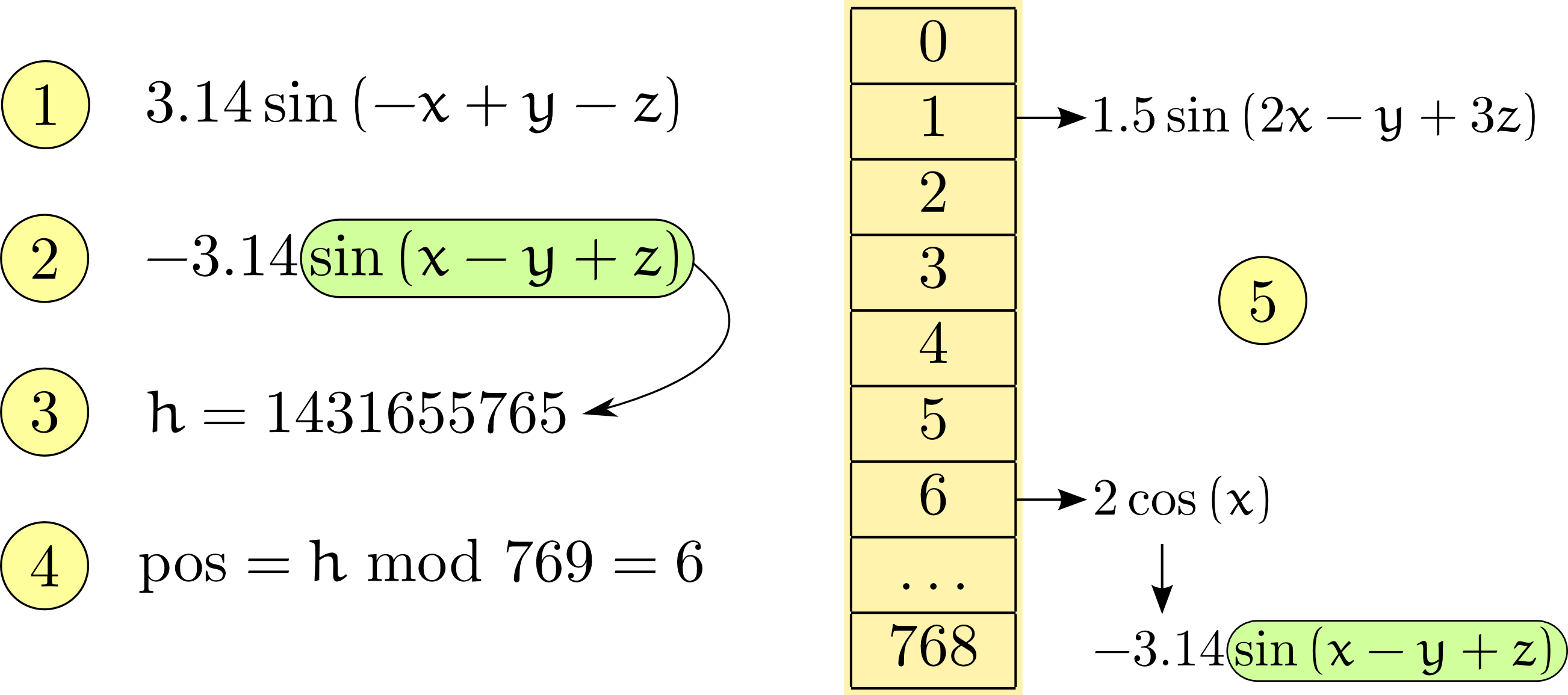}
\par\end{centering}

\caption{Insertion of a term into a Fourier series. The term (1) is first canonicalised
(2), then its hash value, $h$, is calculated from the key (3). The
position in the hash table is calculated through the modulo operation
(4), which, in this example, produces index number 6. Since the sixth
slot of the hash table contains a term with a different key, the term being inserted is then
chained in a linked list (5).\label{fig:insert_01}}

\end{figure}

The single point of entry for terms in the base series class is the
\lstinline[basicstyle={\ttfamily}]!insert()! method. This method
is a wrapper around \lstinline[basicstyle={\ttfamily}]!hash_set!'s
own \lstinline[basicstyle={\ttfamily}]!insert()! method which performs
additional checks before actually inserting the term into the container.
The most important checks are the following (in order of execution):

\begin{enumerate}
\item term is ignorable: a term is always ignorable when either the coefficient
or the key are mathematically equivalent to zero. Additionally, coefficients
and keys may implement additional ignorability criterions (double-precision
coefficients, for instance, are considered ignorable when their absolute
value is below a critical numerical threshold). Ignorable terms are
simply discarded;
\item term is in canonical form: the keys of certain series types might
have multiple valid mathematical representations. The two trigonometric
keys\begin{eqnarray*}
 & \cos\left(x-y\right),\\
 & \cos\left(-x+y\right),\end{eqnarray*}
for instance, are mathematically equivalent, yet their straightforward
translations into an array-like container will be different:\begin{eqnarray*}
 & \left[1,-1,\textnormal{true}\right],\\
 & \left[-1,1,\textnormal{true}\right].\end{eqnarray*}
To remove such ambiguity, trigonometric keys, if necessary, are reduced
during insertion to a unique canonical representation in which the
first integer element is always positive (in case of sine trigonometric
keys the canonicalisation of the key must be followed by a change
in the sign of the coefficient);
\item term is not unique: before insertion of a non-ignorable term in canonical
form it is checked whether an equivalent term exists in the series.
This operation requires a lookup operation in the \lstinline[basicstyle={\ttfamily}]!hash_set!.
If an equivalent term exists, the coefficient of the term being inserted
is added to the existing equivalent term. Otherwise, the term is inserted
as a new element.
\end{enumerate}
The base series' \lstinline[basicstyle={\ttfamily}]!insert()! method
is used when loading series from files, when performing addition/subtraction
of series (see \S\ref{sub:Addition-and-subtraction}) and in general
whenever it is necessary to manipulate directly the terms of the series
(such as in the expansions described in \S\ref{sec:Nontrivial-operations-on}).
Figure \ref{fig:insert_01} visualises the insertion of a Fourier
series term into Boost's \lstinline[basicstyle={\ttfamily}]!hash_set!
data structure.%

\section{Series arithmetic}

Beside being used frequently in actual computations, arithmetic on
series constitutes also the basis of more advanced manipulations (such
as the series expansions described in \S\ref{sec:Nontrivial-operations-on}).
An efficient implementation of basic arithmetic is then a fundamental
component of an efficient specialised manipulator.

The sets of series types encountered in Celestial Mechanics typically
constitute Abelian groups under the operations of addition and subtraction
whose identity elements are series with zero terms. Additionally,
such sets are closed under the operation of multiplication, but, in
general, it will not possible to express inverse series within a given
set in finite terms. In the context of Celestial Mechanics, an effective
way to compute real powers (and hence also the inversion) of certain families of series is the generalised
binomial theorem, as described in \S\ref{sec:Nontrivial-operations-on}.

\subsection{Addition and subtraction\label{sub:Addition-and-subtraction}}

The adoption of hashing techniques, as described in \S\ref{sec:Storage-of-terms},
allows for a straightforward implementation of the operations of series
addition and subtraction. To add a series $S_{2}$ to a series $S_{1}$,
it will be enough to simply \lstinline[basicstyle={\ttfamily}]!insert()!
all terms belonging to $S_{2}$ into $S_{1}$. As described earlier,
the insertion routine will take care of dealing with equivalent terms;
in case of a subtraction, the terms' signs will be changed before
insertion.

Since hash tables feature expected $\OO{1}$ complexity for element
insertion, series addition and subtraction will feature expected $\OO{n}$
complexity, where $n$ is the number of terms of the series being
added or subtracted.

\subsection{Multiplication}

Series multiplication is the most time-consuming among the basic operations
that can be performed on series, and hence an obvious target for heavy
optimisation. Generally speaking, to multiply two series $S_{1}$
and $S_{2}$ consisting of $n_{1}$ and $n_{2}$ terms respectively,
it will be necessary to multiply each term of $S_{1}$ by each term
of $S_{2}$. It follows then that it will be needed to perform $n_{1}n_{2}$
term-by-term multiplications and insertions into the resulting series.
The complexity of series multiplication is hence quadratic, or $\OO{n^{2}}$,
with respect to term-by-term multiplications and insertion operations.

We recall here that polynomials can be multiplied by using asymptotically-faster
algorithms, like Karatsuba (\cite{Karatsuba}) and FFT algorithms
(\cite{BrighamFFT}). Such algorithms however are not usually employed
in Celestial Mechanics for the following reasons:

\begin{itemize}
\item they work best on the assumption that the polynomials to be multiplied
are dense (\cite{Fateman03}). This is not generally the case in Celestial
Mechanics;
\item these algortihms are practically faster than $\OO{n^{2}}$ multiplication
for polynomial degrees that are too high to be encountered in typical
Celestial Mechanics problems.
\end{itemize}
We are not currently aware of algebraic manipulators specialised for
Celestial Mechanics using fast polynomial multiplication algorithms.

We can implement multiplication similiarly to addition and subtraction,
i.e., by performing term-by-term multiplications and accumulating
the results into a newly-created series using the \lstinline[basicstyle={\ttfamily}]!insert()!
method. Indeed, that's precisely what happens in Piranha for series
in nonzero echelon levels, i.e., when dealing with coefficient
series. In such cases the time needed for term-by-term multiplication
dwarfs the time needed for term insertion, so it is of little benefit
to optimise the performance of term accumulation.

Echelon-0 series, however, feature numerical coefficients, and hence the cost of term multiplication might become
comparable to or greater than the cost of term insertion (especially with double-precision
coefficients). In this situation the cost of memory allocations and
the poor spatial locality exhibited by hash tables employing separate
chaining can become performance bottlenecks. For this reason in Piranha
we adopt an intermediate representation\footnote{Other specialised manipulators employ intermediate representations
during series multiplication. The specialised manipulator TRIP, for instance, uses burst tries
(\cite{Heinz2002}) for the intermediate storage of series (\cite{gastineau2006}).} for the multiplication of suitable echelon-0 series based on an procedure
that we conceived autonomously, before learning that it is essentially
an application of a known algorithm named \emph{Kronecker substitution}
(\cite{kronecker82}, \cite{Moenck76}).

\subsubsection{Kronecker substitution}

Kronecker substitution can be intuitively understood in the case of
multivariate polynomials with the aid of a simple table laying out the monomials
in lexicographic order. The lexicographic representation for three
variables $x$, $y$ and $z$, and up to the third power in each variable
is displayed in Table \ref{tab:Kronecker-substitution-for}.%
\begin{table}
\begin{centering}
\begin{tabular}{cccc}
\toprule$z$ & $y$ & $x$ & Code\tabularnewline
\midrule0 & 0 & 0 & 0\tabularnewline
0 & 0 & 1 & 1\tabularnewline
0 & 0 & 2 & 2\tabularnewline
0 & 0 & 3 & 3\tabularnewline
\midrule0 & 1 & 0 & 4\tabularnewline
0 & 1 & 1 & 5\tabularnewline
0 & 1 & 2 & 6\tabularnewline
0 & 1 & 3 & 7\tabularnewline
\midrule0 & 2 & 0 & 8\tabularnewline
$\ldots$ & $\ldots$ & $\ldots$ & $\ldots$\tabularnewline
3 & 3 & 3 & 63\\ \bottomrule\tabularnewline
\end{tabular}
\par\end{centering}

\caption{Kronecker substitution for a 3-variate polynomial up to the third
power in each variable.\label{tab:Kronecker-substitution-for}}

\end{table}
The column of codes is obtained by a simple enumeration of the exponents'
multiindices. It can be noted how in certain cases the addition of
multiindices (and hence the multiplication of the corresponding monomials)
maps to the addition of their codified representation. For instance:\begin{equation}
c\left(\left[0,0,3\right]\right)+c\left(\left[0,1,0\right]\right)=3+4=7=c\left(\left[0,1,3\right]\right)=c\left(\left[0,0,3\right]+\left[0,1,0\right]\right),\label{eq:c_01}\end{equation}
where we have noted with $c=c\left(\mathbf{M}\right)$ the function
that codifies a multiindex vector $\mathbf{M}$. An inspection of
Table \ref{tab:Kronecker-substitution-for} promptly suggests that
for an $m$-variate polynomial up to the $n$-th power, $c$'s effect
is equivalent to a scalar product between the multiindices vectors
and a constant coding vector $\mathbf{c}$ defined as\begin{equation}
\mathbf{c}=\left[\left(n+1\right)^{m-1},\left(n+1\right)^{m-2},\ldots,n+1,1\right],\end{equation}
so that eq. (\ref{eq:c_01}) can be generalised as\begin{equation}
\mathbf{c}\cdot\left(\mathbf{M}_{1}+\mathbf{M}_{2}\right)=\mathbf{c}\cdot\mathbf{M}_{1}+\mathbf{c}\cdot\mathbf{M}_{2}.\label{eq:cod_vector_01}\end{equation}
While this equation is valid in general due to the distributivity
of scalar multiplication, the codification of a single multiindex
will produce a unique code only if the multiindex is representable
within the representation defined by $\mathbf{c}$. For the codification
represented in Table \ref{tab:Kronecker-substitution-for}, for instance,
we have\begin{equation}
\mathbf{c}=\left[\left(3+1\right)^{2},3+1,1\right]=\left[16,4,1\right].\label{eq:c33}\end{equation}
If we try to code a monomial in which at least one of the exponents
is greater than 3 using eq. (\ref{eq:c33}), we can see how the mapping
between multiindices and codes is not univocal any more. For instance
the following different multiindices will produce the same code:

\begin{eqnarray}
\left[0,0,4\right]\cdot\left[16,4,1\right] & = & 4,\\
\left[0,1,0\right]\cdot\left[16,4,1\right] & = & 4.\end{eqnarray}

Kronecker substitution constitutes an addition preserving homomorphism
between the space of integer vectors whose elements are bound in a
range and a subset of integers. It is equivalent to considering the
elements of the multiindex as the digits of a number in base $n+1$,
and transforming it into its decimal counterpart. In the 3-variate
example above, for instance, the last row of the table, 333, is decimal
63 in base 4. It also provides a method to reduce multivariate polynomials
into univariate ones, as the codes can be seen as exponents of a univariate
polynomial.

In Piranha we have adopted a generalisation of Kronecker substitution
in which

\begin{itemize}
\item we consider variable codification, i.e., each element of the multiindex
has its own range of variability,
\item we extend the validity of the codification to negative integers and
subtraction operations.
\end{itemize}
The first generalisation allows to compact the range of the codes.
If, for instance, the exponent of variable $x$ in the example above
varies only from 0 to 1 (instead of varying from 0 to 3 like for $y$
and $z$), we can avoid codes that we know in advance will be associated
to nonexisting monomials (i.e., all those in which $x$'s exponent
is either 2 or 3). The second generalisation, which derives from the
fact that under an appropriate coding vector it is possible to change
$\mathbf{M}_{2}$'s sign in eq. (\ref{eq:cod_vector_01}) retaining
the validity of the homomorphism, allows to apply Kronecker substitution
to the multiplication of Fourier series. Fourier series multiplication,
indeed, requires the ability not only to add but also to subtract
the vectors of trigonometric multipliers according to the elementary
trigonometric formulas\begin{equation}
\begin{gathered}A\cos\alpha\cdot B\cos\beta=\frac{AB}{2}\cos\left(\alpha-\beta\right)+\frac{AB}{2}\cos\left(\alpha+\beta\right),\\
A\cos\alpha\cdot B\sin\beta=\frac{AB}{2}\sin\left(\alpha+\beta\right)-\frac{AB}{2}\sin\left(\alpha-\beta\right),\\
A\sin\alpha\cdot B\cos\beta=\frac{AB}{2}\sin\left(\alpha-\beta\right)+\frac{AB}{2}\sin\left(\alpha+\beta\right),\\
A\sin\alpha\cdot B\sin\beta=\frac{AB}{2}\cos\left(\alpha-\beta\right)-\frac{AB}{2}\cos\left(\alpha+\beta\right).\end{gathered}
\label{eq:werner}\end{equation}
As far as we have been able to verify, we have not found any indication
of previous uses of Kronecker substitution to perform multiplication
of Fourier series.

Table \ref{tab:Generalised-Kronecker-codification} shows the generalised
Kronecker substitution for a multivariate polynomial (or Fourier series)
in which the exponents (or trigonometric multipliers) vary on different
ranges.%
\begin{table}
\begin{centering}
\begin{tabular}{crccc}
\toprule$x_{m-1}$ & $\ldots$ & $x_{1}$ & $x_{0}$ & Code\tabularnewline
\midrule$e_{m-1,\min}$ & $\ldots$ & $e_{1,\min}$ & $e_{0,\min}$ & $0$\tabularnewline
$e_{m-1,\min}$ & $\ldots$ & $e_{1,\min}$ & $1+e_{0,\min}$ & $1$\tabularnewline
$\ldots$ & $\ldots$ & $\ldots$ & $\ldots$ & $\ldots$\tabularnewline
$e_{m-1,\min}$ & $\ldots$ & $e_{1,\min}$ & $e_{0,\max}$ & $e_{0,\max}-e_{0,\min}$\tabularnewline
\midrule$e_{m-1,\min}$ & $\ldots$ & $1+e_{1,\min}$ & $e_{0,\min}$ & $1+e_{0,\max}-e_{0,\min}$\tabularnewline
$e_{m-1,\min}$ & $\ldots$ & $1+e_{1,\min}$ & $1+e_{0,\min}$ & $2+e_{0,\max}-e_{0,\min}$\tabularnewline
$e_{m-1,\min}$ & $\ldots$ & $1+e_{1,\min}$ & $2+e_{0,\min}$ & $3+e_{0,\max}-e_{0,\min}$\tabularnewline
$\ldots$ & $\ldots$ & $\ldots$ & $\ldots$ & $\ldots$\\ \bottomrule\tabularnewline
\end{tabular}
\par\end{centering}

\caption{Generalised Kronecker substitution for an $m$-variate polynomial
(or Fourier series) in which each exponent (or trigonometric multiplier)
varies on a different range.\label{tab:Generalised-Kronecker-codification}}

\end{table}
 If we define\begin{eqnarray}
\mathbf{e} & = & \left(e_{0},e_{1},\ldots,e_{m-1}\right),\\
\mathbf{e}_{\min/\max} & = & \left(e_{0,\min/\max},e_{1,\min/\max},\ldots,e_{m-1,\min/\max}\right),\\
c_{k} & = & 1+e_{k,\max}-e_{k,\min},\\
\mathbf{c} & = & \left(1,c_{0},c_{0}c_{1},c_{0}c_{1}c_{2},\ldots,\Pi_{k=0}^{m-2}c_{k}\right),\\
\chi & = & \mathbf{c}\cdot\mathbf{e}_{\min},\end{eqnarray}
it is easy to show that the code of the generic multiindex $\mathbf{e}$
is obtained by\begin{equation}
c\left(\mathbf{e}\right)=\mathbf{c}\cdot\mathbf{e}-\chi.\end{equation}

Codifying monomials and trigonometric keys into integers using Kronecker
substitution yields some advantages. First, just like in traditional integer packing
techniques, we can save space by using a single integer to represent
many integers. Second, we can map both multiindex addition and subtraction
to the corresponding operation on a single integer. Finally, in the
context of hashing techniques, the codes obtained through Kronecker
substitution constitute \emph{perfect} hash values, in the sense that
the mapping between the multiindices and the corresponding codes is
univocal. Then, depending on the range of the codes and on the characteristics
of the input series, we can adopt two strategies to perform term accumulation
during series multiplication:

\begin{enumerate}
\item \emph{perfect hashing}: if there is enough memory available and the
input series are not much too sparse, then it is convenient to use
the codes of the resulting terms directly as indices in an array (i.e.,
in a perfect hash table). In this way to accumulate series terms during
multiplication it will be enough to multiply term-by-term the coefficients,
add/subtract the codes and write directly in the memory address indicated
by the newly-generated code;
\item \emph{sparse hashing}: if the input series are much too sparse or
there is not enough memory, then the resulting codes can be reduced
through the modulo operation to a smaller value and used to place
the terms resulting from the multiplication in a standard (i.e., non-perfect)
hash table.
\end{enumerate}
Both these algorithms are discussed in greater detail in \S\ref{sub:Cache-friendly-perfect-and}.

In order for Kronecker substitution to be effective, it is necessary
that the codes vary within a range representable with hardware integers,
otherwise the speed benefits related to the codification are reduced.
On modern 64 bits architectures, this restriction is seldom limiting
in the context of Celestial Mechanics. The monomials of a 6-variate
polynomial, for instance, can be codified with Kronecker substitution
into 64 bits unsigned integers up to exponent 1624 in all variables. In any
case, if Kronecker substitution is not feasible, Piranha will revert
to the algorithm, described above, used for nonzero echelon level
series.

The main disadvantage of using Kronecker substitution is the need
to decode the resulting series. Depending on the density of the input
series, the time needed to perform this step can become a relevant
fraction of the total time needed for series multiplication.

\subsubsection{Cache-friendly perfect and sparse hashing on codified series\label{sub:Cache-friendly-perfect-and}}

Both the perfect and sparse hash algorithms mentioned above perform
the following steps:

\begin{enumerate}
\item code the input series into vectors of nonzero coefficient-code pairs
using Kronecker substitution;
\item multiply coefficient by coefficient the two series, using the addition
(and subtraction, for Fourier series) of codes to establish the placement
of each resulting coefficient in an appropriate data structure;
\item decode the result into the input series type.
\end{enumerate}
In the case of perfect hashing, the data structure used to store the
result of coded series multiplication is a simple array of coefficients.
The index of each coefficient resulting from each term-by-term multiplication
will simply be given by its corresponding code, resulting from the
addition or subtraction of the input codes. In perfect hashing, hence,
each term-by-term multiplication consists of a coefficient multiplication,
one or two integer additions and one or two memory redirections.

We have to resort to sparse hashing when we cannot use an array to
store the resulting coefficients, because an array would be either
too large to fit in the available memory or too costly to allocate
and setup because of the high sparseness of the input series (which
usually translates in high sparseness also for the resulting series).
Thus, for sparse hashing we have implemented a cache-friendly hash
table for the storage of coefficient-code pairs that minimises memory
allocations and tries to maximise spatial locality of reference. Although
different in implementation, the logic of term lookup and insertion
is the same as explained in Figure \ref{fig:insert_01}. Our implementation
is essentially a hash table employing separate chaining in which the
buckets have fixed sizes and are laid out contiguously in a single
dynamically-allocated memory area. In such a structure a rehash is
triggered not when the load factor reaches a certain threshold (as
it usually happens in standard hash table implementations) but when
a bucket is completely filled up. To reduce the need for rehashing,
a number of buckets is used as an overflow area to store coefficient-code
pairs belonging to filled-up buckets, so that the rehash is dealyed
until the overflow buckets are exhausted. The overflow buckets must
be checked at every lookup operation. According to our tests, this
hash table variant with a bucket depth of 12 is usually filled up
to $\sim30\%$ before needing a rehash.

The design of the hash table used for sparse hashing aims at maximising
performance by utilising efficiently the cache memory found on modern
computer architectures. The buckets, for instance, consist of arrays
instead of linked lists (as commonly implemented in separate chaining)
in order to improve spatial locality of reference, so that the load
of the first element of the bucket from the slow RAM into the fast
cache memory entails also the loading of a number of successive elements.
If a linked list were used instead, such prefetching of successive
elements would not be possible since in linked lists the second and
successive elements are generally not contiguous to the first one;
buckets implemented as linked lists hence require a load operation
from RAM for every element.

We adopted two other techniques in order to optimise cache memory
utilisation for both perfect and sparse hashing:

\begin{enumerate}
\item ordering of coded input series: in both algorithms the codes resulting
from Kronecker substitution are used to calculate a memory address
(they are used directly as an index in perfect hashing, while in sparse
hashing they are used as an index after the application of the modulo
operation). If we order the coded input series according to the code
and we proceed to term-by-term multiplications in two nested \lstinline[basicstyle={\ttfamily}]!for!
cycles, it follows naturally that in each innermost \lstinline[basicstyle={\ttfamily}]!for!
cycle (i.e., while iterating over the terms of the second series having
fixed a term in the first one) we are going to write into successive
memory areas\footnote{While this is always true for polynomials, in the case of Fourier
series there is also a subtraction involved. Thus for Fourier series
this optimisation may be not as effective as for polynomials.}. Doing this both improves prefetching and helps the cache subsystem
determine a predictable memory access pattern. According to our tests,
this optimisation allowed for a reduction in the time needed to multiply
large series up to 40\% with respect to the case in which the input
series are unordered;
\item cache blocking: another way to optimise cache memory utilisation is
to promote temporal locality, i.e., making sure that data is used
as often as possible in order to avoid eviction from cache memory.
This effect can be obtained through blocking techniques: during series
multiplication, instead of fixing a term in the first series and iterating
over all the terms of the second series we iterate just over a portion
of the second series, then moving on to the second term of the first
series and repeating the procedure. In other words, we divide the
series in blocks and multiply block-by-block. This ensures that we
are going to read and write the same memory areas very frequently,
thus maximising the permanence in cache memory. In our tests this
optimisation can lead to substantial speed gains: in the case of large
polynomials the reduction of the time needed to perform multiplications
can reach 70\%.
\end{enumerate}
The version of Piranha benchmarked in \S\ref{sec:Benchmarks} employs
the optimisation techniques described in this section.

\section{Nontrivial operations on series\label{sec:Nontrivial-operations-on}}

In Celestial Mechanics it is common to encounter algebraic expressions
that cannot be mapped directly to Poisson series or polynomials. An
expression often arising is, for instance, the square root\begin{equation}
\sqrt{1-e^{2}},\label{eq:root_01}\end{equation}
where $e$ is an orbital eccentricity. In other cases it may be necessary
to compute the sine or cosine of a Poisson series.

Such occurrences are usually dealt with through series expansions.
Expression (\ref{eq:root_01}) may be rewritten in terms of the MacLaurin
development\begin{equation}
\sum_{n=0}^{\infty}\frac{\left(2n\right)!}{\left(1-2n\right)n!^{2}4^{n}}e^{2n}\label{eq:maclaurin_root_01}\end{equation}
for $\left|e\right|<1$. Such an expansion truncated to a finite order
is often appropriate since in most practical problems of perturbation
theory the value of $e$ is small. In Piranha we provide two general
series expansions that can be useful to reduce to a representable
algebraic form many common occurrencies of problematic expressions.

The first series expansion is employed to compute the real power of
a series and uses the generalised binomial theorem (\cite{Arfken85}):\begin{equation}
\left(x+y\right)^{r}=\sum_{k=0}^{\infty}{r \choose k}x^{r-k}y^{k},\end{equation}
where, for our purposes, $r\in\Re$ and $x$ and $y$ are complex
numbers. In Piranha the real power of a series $S$ is rewritten in
terms of a leading term $L$ and a tail series $T$, so that\begin{eqnarray}
\left(S\right)^{r} & = & \left(L+T\right)^{r}\\
 & = & \sum_{k=0}^{\infty}{r \choose k}L^{r-k}T^{k}.\end{eqnarray}
It is the responsibility of the user to make sure that the series
is convergent (i.e., $r$ is a positive integer or $\left|L\right|>\left|T\right|$).
Piranha will choose the leading term adopting series-specific criterions:
for Poisson series and polynomials, for instance, the leading term
is the one with lowest total degree, while for Fourier series the
leading term is the one with the highest coefficient in absolute value.
It is worth noting that while $T$ is subject only to trivially-implementable
natural powers, $L$ must be in general amenable to real exponentiation.
In case of Poisson series, the real power of the leading term $L$
involves the calculation of the real power of its polynomial coefficient,
so that the binomial expansion may be called again to compute $L^{r-k}$
in a somewhat recursive fashion. This method of calculation of real
powers can be succesfully applied to expressions such as (\ref{eq:root_01})
(effectively producing the same expression as (\ref{eq:maclaurin_root_01}))
or to compute the inverse of a suitable series.

The other series expansion implemented in Piranha is used to compute
the complex exponential of real Poisson and Fourier series, and it
is based upon the Jacobi-Anger expansion (\cite{WatsonTTB}):\begin{equation}
\exp\left[\imath x\cos\theta\right]=\sum_{n=-\infty}^{\infty}\imath^{n}J_{n}\left(x\right)\exp\left[\imath n\theta\right],\label{eq:jacang_01}\end{equation}
where $J_{n}\left(x\right)$ is the Bessel function of the first kind
of integer order $n$, which can be expressed by the MacLaurin development

\begin{equation}
J_{n}(x)=\sum_{l=0}^{\infty}\frac{(-1)^{l}}{2^{2l+n}l!(n+l)!}x^{2l+n}.\label{eq:bessel_01}\end{equation}
By using eq. (\ref{eq:jacang_01}) and eq. (\ref{eq:bessel_01}),
and by remembering the property of Bessel functions\begin{equation}
J_{-n}\left(x\right)=\left(-1\right)^{n}J_{n}\left(x\right),\end{equation}
it is possible to transform the complex exponential of a Poisson series
into a product of Poisson series. From the complex exponential, the
cosine and sine of the Poisson series can be extracted as the real
and imaginary parts respectively. As far as we were able to verify
in the literature, no other manipulator implements circular functions
of Poisson series using the Jacobi-Anger expansion. Usually such calculations
are performed by decomposing the Poisson series into a leading term
and a tail (similarly to what is done in the binomial expansion) and
by expanding the function through elementary trigonometric formulas:\begin{eqnarray}
\cos S & = & \cos\left(L+T\right)\\
 & = & \cos L\cos T-\sin L\sin T.\end{eqnarray}
At this point $\cos T$ and $\sin T$ are expressed through truncated
MacLaurin series. Another approach, recently proposed in \cite{martinez07}
and based on a Richardson-type elimination scheme, has been shown
to be advantageous in certain cases with respect to Taylor expansions
both in terms of accuracy and computing time.

With the ability to compute real exponentiation and circular functions
on series, it is possible to implement an array of capabilities consisting
of special functions and series expansions relevant to Celestial Mechanics
applications (thus implementing the basis of what is sometimes called a \emph{keplerian
processor} - see \cite{Broucke70} and \cite{Brumberg89}). Piranha currently implements
Bessel functions of the first kind, Legendre polynomials, associated
Legendre functions, (rotated) spherical harmonics and the elliptic
expansions for $\frac{r}{a}$, $\frac{a}{r}$, $E$, $f$, $\begin{array}{c}
\cos\\
\sin\end{array}\left(f\right)$ and $\exp\left[\imath nE\right]$ (as seen in standard Celestial
Mechanics textbooks, such as \cite{SSD2000}, Chapter 2).

\section{Python bindings}

Piranha's core is a C++ library that can be used within any C++ program.
Although we have tried to make the use of the library as friendly
as possible (e.g., by using operator overloading), usage as a C++
library remains somehow uncomfortable: a basic knowledge of the C++
programming language is required, and every change of the source code
in the main routine must be followed by a recompilation of the whole
program, a task that can become slow and cumbersome in the case of
generically-programmed libraries like Piranha. Additionally, direct
usage as a C++ library is not interactive.

For these reasons we provide a set of Python bindings (called Pyranha)
which is intended to be the preferred way of interacting with the
manipulator for end-users. In Pyranha, Piranha's C++ classes are exposed as Python
objects with the help of the Boost.Python library. It is important
to stress that this is not a translation of C++ code into Python:
even as Python objects, Piranha's classes are still compiled C++ code,
so that there is no performance penalty in using the manipulator from
Python.

Using Piranha from an interpreted language like Python has several
advantages. In our opinion, two important ones are the following:

\begin{enumerate}
\item Python (\cite{Python}) is an open-source and widely used language,
with a standardised syntax, an easy learning curve and a vast amount
of publicly-available additional modules;
\item Python can be used interactively. Through enhanced Python shells and
graphics libraries it is possible to provide an interactive environment
with graphical capabilities. Pyranha takes advantage of the features
provided by the IPython shell and the matplotlib graphic library (\cite{IPython,Matplotlib}).
\end{enumerate}
Among Python's features, of particular relevance for Pyranha are lambda
functions, which allow to define small functions inline directly when
they are called. Lambda functions are particularly handy for two useful
series methods available in Pyranha. The first one is a plotting method,
which works in conjunction with the matplotlib module. This method
takes as parameter a function that specifies what to plot. The following
Python code snippet uses a lambda function to specify that it is being
requested to plot the norms of the coefficients of the series \lstinline[basicstyle={\ttfamily},language=Python]!foo!:
\begingroup
\inputencoding{latin1}
\begin{lstlisting}[basicstyle={\ttfamily},language=Python]
foo.plot(key = lambda t: t.cf.norm)
\end{lstlisting}
\endgroup
This means that from every term \lstinline[basicstyle={\ttfamily},language=Python]!t!
of the series, the coefficient's norm is extracted and plotted. Another
useful method is the \lstinline[basicstyle={\ttfamily},language=Python]!filter()!
method, which is modelled after Python's builtin function with the
same name. \lstinline[basicstyle={\ttfamily},language=Python]!filter()!
is used to extract from series terms satisfying a criterion specified
through a lambda function:
\begingroup
\inputencoding{latin1}
\begin{lstlisting}[basicstyle={\ttfamily},language=Python]
foo.filter(lambda t: t.key.freq > 0)
\end{lstlisting}
\endgroup
In this case we are extracting all terms of a Poisson/Fourier series
whose frequency (which is a property of the trigonometric key) is
positive. \lstinline[basicstyle={\ttfamily},language=Python]!filter()!
can work in a recursive fashion in case of nonzero echelon level series.
This means that it is possible to pass more than one argument: the
first argument is used to filter the terms of the series, the second
argument to filter the terms of each coefficient series in the terms
surviving from the first filtering, and so on. \lstinline[basicstyle={\ttfamily},language=Python]!filter()!
can be useful in perturbation theories, when it is necessary to select
the terms of the series which are relevant to the physical problem
that is being solved.

\section{Benchmarks\label{sec:Benchmarks}}

In this section we present a few benchmarks on the multiplication
of echelon-0 series. The systems benchmarked are: SDMP, a library
for sparse polynomial multiplication and division (\cite{monagan07}),
TRIP, an algebraic manipulator devoted to Celestial Mechanics (\cite{TRIP2008}),
Pari, a computer algebra system for number theory (\cite{PARI2}),
Magma (\cite{bosma97}), Singular (\cite{Singular}) and Maple. The
usual caveats about benchmarking as a method to measure performance
apply. Additionally, we must preface the following disclaimers:

\begin{itemize}
\item measurements for other systems were taken by SDMP co-author Roman
Pearce on a Linux server equipped with a Core2 Xeon 3.0 GHz CPU with
4MB of L2 cache, while Piranha's measurements were taken on a Linux
laptop equipped with a Core2 1.8 GHz CPU with 2MB of L2 cache (the
GCC compiler version 4.3.2 was used to compile Piranha). We have chosen
to express the results as both the original timings in seconds and
clock cycles per term-by-term multiplication (ccpm) in an effort to
provide a uniform scale. It must be noted however that the latter
unit of measure fails to give meaningful results when the computation
time is bounded by memory transfers (i.e., when most of the time is
spent loading data from RAM). In such cases the speed of execution
of the same task will depend more on the speed of the memory subsystem
than on the CPU speed;
\item we have chosen to express Piranha's results as sums of two numbers:
the first number measures the speed of the actual multiplication,
while the second number represents the time spent unpacking the
result from the coded series representation back into the standard
series representation (as explained at the beginning of \S\ref{sub:Cache-friendly-perfect-and}).
The reason for this distinction will be explained in the comments
about the individual benchmarks;
\item the computer algebra systems tested adopt different representations
for coefficients and exponents (or trigonometric multipliers), so the
relevance of these benchmarks is related to the adequacy of the representations
used in the various systems with respect to the task at hand. Additionally,
there are differences also in the representation of series (certain
systems, such as Pari and TRIP for instance, use a recursive representation
for polynomials, while SDMP's and Piranha's representations are distributed).
\end{itemize}
We would like to gratefully thank Roman Pearce for allowing us to
reproduce the benchmarks he performed. The original benchmarks are
available online at the following address:

\begin{center}
\url{http://www.cecm.sfu.ca/~rpearcea/}
\par\end{center}

\subsection{Fateman's benchmark}

The first benchmark is modelled after that proposed by Richard
Fateman in \cite{Fateman03}, and consists of the calculation of\begin{equation}
s\left(s+1\right),\end{equation}
where $s$ is defined as the multivariate polynomial\begin{equation}
s=\left(1+x+y+z+t\right)^{30}.\end{equation}
The computation is expressed as $s\left(s+1\right)$ instead of, e.g.,
$s^{2}+s$, to prevent ``smart'' computer algebra systems to calculate
$s^{2}$ more efficiently than through a multiplication (e.g., by
multinomial expansion). $s$ and $s+1$ consist of $46\,376$ terms,
while the resulting polynomial consists of $635\,376$ terms. The
results are displayed in Table \ref{tab:fateman}. Some necessary remarks:

\begin{table}
\begin{center}
\begin{tabular}{lr}
\toprule System & Timings (seconds, ccpm)\tabularnewline
\midrule Piranha 2008.11 (double-precision) & 5.4 + 0.6, 4.5 + 0.5\tabularnewline
SDMP September 2008 (monomial = 1 word) & 47, 65\tabularnewline
TRIP v0.99 (double-precision) & 44, 61\tabularnewline
Pari 2.3.3 (w/ GMP) & 512, 714\tabularnewline
Magma V2.14-7 & 679, 947\tabularnewline
Singular 3-0-4 & 1482, 2067\tabularnewline
Maple 11 & 15986, 22298\\ \bottomrule\tabularnewline
\end{tabular}
\end{center}
\caption{Timings for Fateman's benchmark.\label{tab:fateman}}
\end{table}

\begin{itemize}
\item the densities of the input polynomials are high enough to make the
perfect hashing algorithm viable;
\item double-precision coefficients, used by Piranha and TRIP, do not allow
to represent exactly the coefficients of the resulting polynomial;
\item SDMP uses inline assembly for the multiplication of 61 bits input
integer coefficients into 128 bits integer coefficients. On Core2
CPUs, this is expected to cost around 4 times the cost of double-precision
multiplication;
\item other systems use arbitrary-precision coefficients.
\end{itemize}
In the case of manipulators devoted to Celestial Mechanics, this test
represents more of a limit case than a computation likely to be encountered
in actual applications.

In the case of Piranha's timing we remark that, since double-precision
multiplication on Core2 CPUs costs around 4 clock cycles when working
in cache, the use of the perfect hashing algorithm allows to approach
the theoretical speed limit for this kind of computation performed
with classical algorithms. When disabling the cache optimisations
described in \S\ref{sub:Cache-friendly-perfect-and} the running time
for this benchmark is around 36 seconds. We also note how the time
needed to unpack the resulting coded polynomial (0.6 s) is a small
fraction of the time needed for the multiplication of the input polynomials.
Finally, for comparison we note that if Piranha is benchmarked using
multiprecision GMP integer coefficients, the running time for this
benchmark becomes 90 s (or 75 ccpm).

\subsection{Sparse polynomial multiplication benchmark\label{sub:Sparse-polynomial-multiplication}}

This benchmark was proposed originally by Michael Monagan and Roman
Pearce, and consists of the following polynomial multiplication:\begin{equation}
\left(1+x+y+2z^{2}+3t^{3}+5u^{5}\right)^{12}\cdot\left(1+u+t+2z^{2}+3y^{3}+5x^{5}\right)^{12}.\end{equation}
Since the polynomials being multiplied consist of $6\,188$ terms
each and the result consists of $5\,821\,335$ terms, this multiplication
is much more sparse than the one featured in the previous benchmark.
For Piranha this means that the perfect hashing algorithm is not viable
and that sparse hashing must be employed instead. The results are
displayed in Table \ref{tab:sparse_polymult}.

\begin{table}
\begin{center}
\begin{tabular}{lr}
\toprule System & Timings (seconds, ccpm)\tabularnewline
\midrule Piranha 2008.11 (double-precision) & 3 + 6.8, 141 + 319\tabularnewline
SDMP September 2008 (monomial = 1 word) & 1.5, 121\tabularnewline
TRIP v0.99 (double-precision) & 1.9, 148\tabularnewline
Pari 2.3.3 (w/ GMP) & 54, 4231\tabularnewline
Magma V2.14-7 & 24, 1880\tabularnewline
Singular 3-0-4 & 59, 4622\tabularnewline
Maple 11 & 333, 26089\\ \bottomrule\tabularnewline
\end{tabular}
\end{center}
\caption{Timings for sparse polynomial multiplication.\label{tab:sparse_polymult}}
\end{table}

The most interesting remark for Piranha is the time spent unpacking
the coded series back into the standard series representation, which
amounts to more than twice the time spent for the actual multiplication.
This behaviour is the combined effect of the following causes:

\begin{itemize}
\item we are using a hash table with standard semantics, which means that
each insertion function will result in a lookup of the term. This
lookup is not necessary because we know that by construction the resulting
coded series features unique terms. By implementing a custom hash
table with an alternative unchecked insertion function we could avoid
pointless term comparisons;
\item to decode the coded series we need to apply repeatedly costly integer
divisions and modulo operations, which are particularly cumbersome
in this benchmark given the high ratio between the number of term-by-term
multiplications and the size of the resulting polynomial. For this
specific benchmark we could avoid codification altogether since there
are no negative exponents and the exponents' values are bounded in
the $\left[0,120\right]$ interval. It follows then that in this case
we could simply pack the exponents as, e.g., 8 bits integers into
a 64 bits word and obtain the same efficiency as Kronecker substitution,
without the need to apply division and modulo operations for the decodification.
This optimisation will be implemented in a future version of Piranha;
\item the \lstinline[basicstyle={\ttfamily}]!hash_set! data structure needs
to perform many memory allocations. While this issue can be mitigated
by using a pool memory allocator, this solution does not avoid frequent
cache-unfriendly memory accesses.
\end{itemize}
As a test, we have tried to see what happens if, instead of unpacking
the series into a \lstinline[basicstyle={\ttfamily}]!hash_set!, we
simply copy the codes into a plain array. This approach is similar
to that adopted by some manipulators which employ plain array-like
storage methodologies. The result is that the total running time is
reduced to 3.2 s, or 150 ccpm. Although this approach would allow
to consistently increase performance, for the time being we are oriented
towards maintaining the current implementation. In addition to the
fact that it is possible to improve performance with the techniques
described above (hash table implementation with additional insertion
semantics, avoiding needless encoding, pool memory allocator), the
current programming model is in our opinion convenient and clear and
allows for capabilities, such as comparison of series in linear time
and identification of terms in constant time, which have proven to
be useful in practice.

\subsection{Fourier series multiplication}

This benchmarks consists of the squaring of the ELP3 Fourier series
representing the solution of the main problem of the lunar theory
for the Moon's distance in the ELP2000 theory (\cite{ELP2000}). The
ELP3 series consists of 702 terms and the squaring produces a series
with $11\,673$ terms\footnote{The exact number of the resulting terms may be subject to variations
in the order of few unities, depending on the value of the term ignorability
threshold.}. The squaring is performed 20 times in order to minimise the effect
of the time needed to load the series from file. The sparseness of
the ELP3 Fourier series, despite higher than in Fateman's benchmark,
is still low enough to make the perfect hashing algorithm viable.

We report in Table \ref{tab:fs_bench} the timings for Piranha and TRIP. We report just the
plain timings because TRIP's results, communicated to us by TRIP's
co-author Micka{\"{e}}l Gastineau, were taken on a CPU which is slower
than the Core2 CPU used for Piranha and whose characteristics we are
not aware of. According to our tests, the Core2 CPU should be $\sim1.5$
times faster.

\begin{table}
\begin{center}
\begin{tabular}{lr}
\toprule System & Timings (seconds)\tabularnewline
\midrule Piranha 2008.11 (double-precision) & 0.495\tabularnewline
TRIP v0.99 (double-precision) & 1.352\\ \bottomrule\tabularnewline
\end{tabular}
\end{center}
\caption{Timings for Fourier series multiplication benchmark.\label{tab:fs_bench}}
\end{table}

In Piranha's case the timings translate in roughly 90 clock cycles
per term-by-term multiplication. It is then evident that here the
overhead of decoding the coded series is dominating, as the multiplication
of Fourier series terms consists of a double-precision multiplication
($\sim4$ clock cycles) and a double-precision division ($\gtrsim6$
clock cycles), as per eqs. (\ref{eq:werner}). To confirm this interpretation
we have computed first $\textnormal{ELP3}^{2}\cdot\textnormal{ELP3}^{2}$
and then $\textnormal{ELP3}^{4}\cdot\textnormal{ELP3}^{4}$ (where
$\textnormal{ELP3}^{4}$ consists of $192\,000$ terms). For these
other two benchmarks we measured a total cost of 15 ccpm and 11 ccpm
respectively, thus confirming that the cache-optimised perfect hashing
algorithm allows to reach a high throughput, especially for long series.

\section{Conclusions, future work and availability}

In this contribution we have provided an overview of the architecture
upon which the Piranha algebraic manipulator is built. We have described
the high-level representation of series, the storage method for terms,
based on hashing techniques, and the algorithms used during series
arithmetic. We have shown how Kronecker substitution can be applied
also to the multiplication of Fourier series and how it allows to
reach promising performance for series multiplication. We also have
briefly described how using Piranha from Python can open up interesting
possibilities of interaction with series objects, and how it is possible
to develop nontrivial functions of series in practical Celestial Mechanics
problems.

Future work on Piranha will target the following areas of interest:

\begin{itemize}
\item implementation of echeloned Poisson series;
\item implementation of multiprecision floating-point coefficients (possibly through
the MPFR library, see \cite{mpfr});
\item implementation of interval arithmetic for both double-precision and
multiprecision coefficients;
\item enhancements for the keplerian processor and implementation of standard algorithms
of perturbation theories in the context of Celestial Mechanics,
\item performance improvements for the multiplication of very sparse series
(as explained in \S\ref{sub:Sparse-polynomial-multiplication});
\item parallelisation for the most time-consuming operations;
\item enhancements for the Python bindings, in order to improve interactivity,
ease of use and flexibility (also by means of graphical user interface
elements).
\end{itemize}
We are also evaluating the possibility to provide some form of interaction
between Piranha and general-purpose computer algebra systems. We are
particularly interested in the possibility to interact with the open-source
mathematics software Sage (\cite{SAGE}).

Piranha is freely available under the terms of the GNU public license,
and it can be downloaded from the website

\begin{center}
\url{http://piranha.tuxfamily.org}
\par\end{center}

The manipulator runs on Unix and Windows platforms, and it can be
compiled with a reasonably recent version of a C++ compiler (the GCC,
Intel and Visual Studio compilers have been successfully used so far).
We hope to be able to gather enough interest around Piranha to establish
a community of users and developers in the future.

\section{Acknowledgements}

We would like to thank Elena Fantino for proofreading this manuscript and for
constant encouragement. We would also like to thank Roman Pearce and Micka{\"{e}}l
Gastineau for testing Piranha and for helpful discussion.

\bibliographystyle{siam}
\bibliography{nemo}

\end{document}